\documentstyle[12pt,epsfig]{article}
\begin{document}
\begin{center}
{\Large \bf{Scale factor duality in quintessence models ?}}
\vspace*{10mm} \\
Elisa Di Pietro\footnote{E-mail: dipietro@astro.ulg.ac.be}  
\vspace*{5mm} \\
Institute of Astrophysics and Geophysics \vspace*{3mm} \\
Theoretical Cosmology Group \vspace*{3mm} \\
University of Li\`ege \vspace*{3mm} \\
B-4000 LIEGE-BELGIUM \vspace*{23mm} \\
\begin{abstract}
\noindent
We consider several kinds of quintessence models in the framework of scale
factor duality. We show that this symmetry exists only for a very small
number of quintessence potentials. We then apply the duality transformations
found to several analytical solutions. It turns out that, in some cases, the
presence of the potential allows a smooth connection between the pre- and
the post-Big Bang phases. This may be a first step toward the resolution
of the singularity {\em problem}.
\end{abstract}
\end{center}
\hspace*{5mm} \vspace*{3mm} 
\newpage
\section{Introduction}
\hspace*{6mm}
{\em Quintessence models} have been proposed as an alternative to the
cosmological constant in order to explain some recent astrophysical
observations which seem to favour a currently accelerating universe
\cite{quintessence}. In these models, a scalar field slowly rolling down its
potential 
is added to the standard cosmological model. Depending on the way it
is introduced, this new field can be interpreted as a fluid with a negative
pressure or as a new ingredient to the gravitation theory. There are several
reasons that lead one to favour a scalar field candidate in place of a
cosmological constant. First of all, while the cosmological constant does
not yet possess a completely satisfactory physical interpretation
\cite{constant}, the scalar field appears naturally in a large number of
alternative theories to general relativity. Furthermore, in some of these
alternative theories, the scalar terms play an important physical role and
consequently cannot be neglected.

The effective string theory is one of those alternative theories. Its study
in the cosmological framework has been motivated by its remarkable property
of symmetry called {\em scale factor duality} (see \cite{duality} and
references therein). This symmetry allows the construction of a {\it dual
solution} from a known solution of the theory field equations. In this
context, the initial solution is used to describe the ``{\em post-Big Bang
phase}'', i.e. the Universe evolution between the initial singularity until
today, whereas the dual solution is associated to a possible ``{\em pre-Big
Bang phase}'', i.e. the Universe as it could have been before the Big Bang.
Though there is no satisfying model explaining the transition between the 
pre- and the post-Big Bang phases, many cosmologists think that
duality symmetry and in particular the pre-Big Bang phase could solve the
singularity {\em problem}. It should be noted, however, that the string
theory is not the only theory having this duality symmetry: many
scalar-tensor theories have this property as well.

As one can see, scale factor duality and quintessence models are both very
promising. However, only quintessence models without a scalar potential have
been studied in the duality framework whereas the supernovae observations
require a model with a potential in order to be interpreted by quintessence.
The purpose of this paper is to see if the scale factor duality can be
maintained in the presence of a scalar potential. The different kinds of
cosmological models
considered are those coming from the scalar-tensor theories with various
couplings between the scalar field, gravitation and the matter source.
For all these models, we determine which form the potential
should have if the theory has to remain invariant under a scale factor
duality symmetry. Finally, we apply the duality transfomations found
to some analytical solutions. 

\section{Non-minimally coupling with gravitation}

The first kind of quintessence models we shall consider is the one coming
from the effective string theory \cite{duality}. These models are the most
studied in the framework of scale factor duality because duality
has initially been developped in the context of the string theory.
The effective string action is given by 
\begin{equation}
S = \frac{1}{2 \kappa}\,\int{ e^{-\phi}\,
\left[ R + \left[ \nabla \phi \right]^2 - V(\phi) \right]\,
\sqrt{- g\,}\,\,d^4 x} + \int{ L_m \sqrt{- g\,}\,\,d^4 x}
\label{action1}
\end{equation}
where $\,\kappa = 8\,\pi\,G\,$ and where a scalar potentiel $V(\phi)$ has
been introduced. In what follows, we shall only consider flat
Friedmann-Lema\^\i tre-Robertson-Walker spacetimes so that the metric tensor
can be written as:
\begin{equation}
\displaystyle
g_{\alpha\,\beta} = \left( 
\begin{tabular}{cc}
$g_{00}(t)$ & $0$ \\
$0$ & $G(t)$
\end{tabular}
\right)
\hspace*{15mm} with \hspace*{15mm}
G(t) = \left(
\begin{tabular}{ccc}
$a^2(t)$ & $0$ & $0$ \\
$0$ & $a^2(t)$ & $0$ \\
$0$ & $0$ & $a^2(t)$ 
\end{tabular}
\right)
\end{equation}
\noindent When the scalar potential is absent, the duality
transformation can be written as \cite{duality} \cite{paper1}:
\begin{equation}
\left\{
\begin{tabular}{rlcl}
$G \,\,\rightarrow$ & $\bar{G}$ & $=$ & $G^{-1}$ \\
$g_{00} \,\,\rightarrow$ & $\bar{g}_{00}$ & $=$ & $g_{00}$ \\
$\phi \,\,\rightarrow$ &
$\bar{\phi}$ & $=$ & $\phi - \,ln \left( det\,G \right)$ \\
$\rho_m \,\,\rightarrow$ &
$\bar{\rho}_m$ & $=$ & $\left[ det\,G \right]\, \rho_m$ \\
$p_m \,\,\rightarrow$ &
$\bar{p}_m$ & $=$ & $- \left[ det\,G \right]\,p_m$
\end{tabular}
\right.
\label{transfo1}
\end{equation}
It remains to see which form the potential should have if we want the
complete
action (\ref {action1}) to remain invariant under duality. We also expect
the potential to be the same function of the scalar field before and after
the duatily transformation. Following the transformation (\ref {transfo1}),
we have $\,e^{-\,\phi}\,\sqrt{-\,g\,}
\,\,\rightarrow\,\, e^{-\,\bar{\phi}}\,\sqrt{- \bar{g}\,}$ which implies
that
the potential must satisfy the following condition:
$\,V(\phi) = \bar{V}(\bar{\phi})\, = V(\bar{\phi})$. This is
possible only if the potential is constant:
\begin{equation}
V(\phi) = \Lambda \,\, \rightarrow \,\, \bar{V}(\bar{\phi}) \,=\, \Lambda
\label{transfo11}
\end{equation}
where $\Lambda$ is a constant. In the framework of the effective string
theory,
this constant can be interpreted as the {\em charge deficit} which is a
constant depending on the compactification scheme used to bring the initial
10 dimensional action to the 4 dimensional action given by (\ref {action1}).
Note that the case of an effective string theory containing a cosmological
constant has already been discussed in the litterature (see, e.g.,
\cite{pot}).

The equation of state for the matter field, i.e. $\,w_m = p_m\,/\,\rho_m\,$,
remains invariant under the scale factor duality only for dust ($w_m = 0$).
Furthermore, it has to be noted that the action (\ref {action1}) with
$V(\phi) = \Lambda$ is equivalent to a Brans-Dicke theory with a linear
potential:
\begin{equation}
\displaystyle 
S = \frac{1}{2\,\kappa} \int{\left(
\Phi\,R + \Phi^{-1}\,\left[\nabla\,\Phi \right]^2 - \Lambda\,\Phi
\right)\,\sqrt{-\,g\,}\,\,d^4 x} 
\label{actionBD}
\end{equation}
where $\Phi = \,e^{-\phi}$, or to a scalar-tensor theory with a
power-law potential:
\begin{equation}
\displaystyle 
S = \frac{1}{2\,\kappa} \int{\left(
\varphi^2\,R + 4\,\left[\nabla\,\varphi \right]^2 -
\Lambda\,\varphi^2 \right)\,\sqrt{-\,g\,}\,\,d^4 x} 
\label{action11}
\end{equation}
where $\varphi = e^{-\,\phi\,/\,2}$. Recently, the scale factor
duality has been discussed in the framework of this scalar-tensor theory
by de Ritis et al. in \cite{deritis}.

In the context of a flat FLRW spacetime ($g_{00} = -1$) filled with dust,
the field equations deriving from the action (\ref {action1})
with $V(\phi) =
\Lambda\,$ are given by
\begin{eqnarray}
\displaystyle
- 6\,\,\frac{\,a'^{\,2}}{a^2} +\,6\,\,\phi'\,\frac{\,a'}{a}
- \phi'^{\,2} + \Lambda
+ 2\,\,e^\phi\,\kappa\,\rho_m & = & 0 \label{eqn11} \\
\displaystyle
2\,\,\phi'' + 4\,\,\phi'\,\frac{\,a'}{a} - \phi'^{\,2}
- 4\,\,\frac{\,a''}{a} - 2\,\,\frac{\,a'^{\,2}}{a^2}
+ \Lambda & = & 0 \label{eqn12} \\
\displaystyle
6\,\,\frac{\,a''}{a} + 6\,\,\frac{\,a'^{\,2}}{a^2}
- 6\,\,\phi'\,\frac{\,a'}{a} - 2\,\,\phi'' + \phi'^{\,2}
- \Lambda & = & 0 \label{eqn13}
\end{eqnarray}
\noindent Meissner and Veneziano have already proposed a solution of this
system when there is no source terms, i.e. when $p_m = \rho_m = 0$
\cite{pot2}. We present here a solution in the presence of the matter
field:
\begin{eqnarray}
a(t) & = & \displaystyle a_0\, e^{\alpha_0\,\tau(t)} \label{sol11}\\
\kappa\,\rho_m(t) & = & \displaystyle A\,a_0^{-3}\,
e^{-\,3\,\alpha_0\,\tau(t)} \\
\phi(t) & = & \displaystyle 3\,\alpha_0\,\tau(t) + ln \left( \,C_1\,
\left[ \,cosh [\sqrt{3\,}\,\alpha_0\,\tau(t) ] - \delta_0 \right]
\right) \label{sol13}
\end{eqnarray}
with the following constraints:
\begin{equation}
\displaystyle
\Lambda = \frac{A^2}{3\,\alpha_0^2\,}\,\left[ 1 - \frac{1}{\delta_0^2}
\right] \hspace*{2cm} 
C_1 = \frac{A\,a_0^3}{3\,\alpha_0^2\,\delta_0} \hspace*{2cm}
\delta_0 \leq 1
\label{constraints}
\end{equation}
The quantity $\tau$ present in (\ref {sol11})-(\ref {sol13}) is called the
{\em conformal time} and is defined by $d \tau = \pm\,a^{-3}\,e^\phi\,dt$.
The function $\tau(t)$ depends on $\delta_0$: 
\begin{eqnarray}
\displaystyle
\tau(t) = \frac{2}{\sqrt{3\,}\,\alpha_0}\,\,arctanh \left[
\pm\,\frac{a_0^3}{\sqrt{3}\,\alpha_0\,C_1\,t} \right]
& if & \delta_0 = 1  \label{case1} \\ & & \nonumber \\
\displaystyle
\tau(t) = \frac{2}{\sqrt{3\,}\,\alpha_0}\,\,arctanh \left(
\frac{\sqrt{1 - \delta_0^2}}{1 + \delta_0}\,tanh \left[ \pm\,
\frac{\sqrt{3\,(1 - \delta_0^2)\,}\,\alpha_0\,C_1\,t}{2\,a_0^3}
\right] \right)
& if & \delta_0 < 1 \label{case2}
\end{eqnarray}
The first relation in (\ref {constraints}) shows that the case given by
(\ref {case1}) is associated to a null cosmological constant whereas the
case given by (\ref {case2}) corresponds to a negative cosmological constant.
Furthermore, the solution without a cosmological constant does not exist for
$\mid t \mid\, < a_0^3\,/\,(\sqrt{3}\,C_1\,\alpha_0)$.

\begin{figure}[h]
\begin{center}
\epsfig{figure=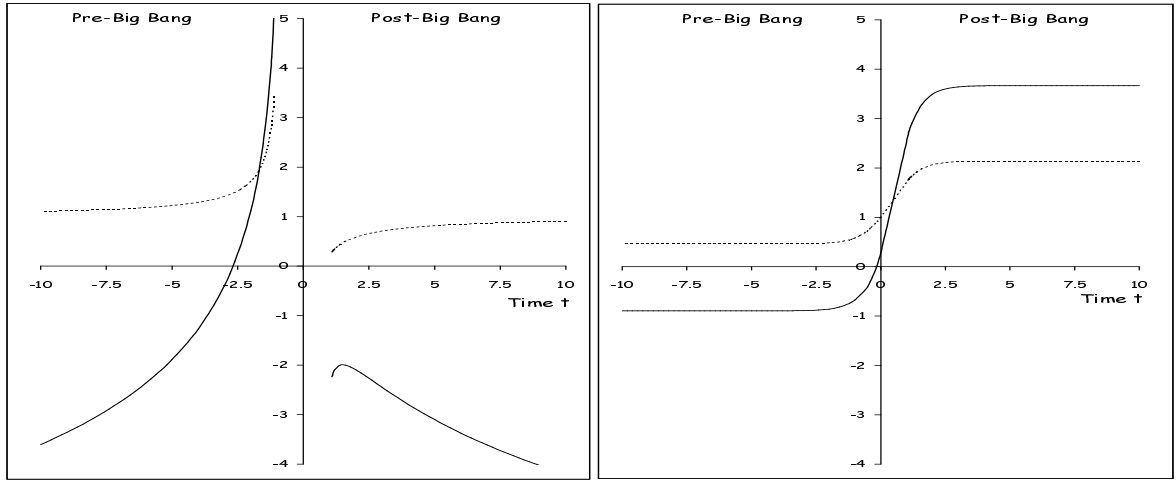,height=65mm,width=16cm}
\end{center}
\caption{
\small{The solution without the cosmogical constant is plotted on
the left graph whereas the one with a negative cosmological constant is
given by the right graph. The dashed line is associated to the scale factor
and the solid line gives the scalar field behavior. For a vanishing
cosmological constant, there is a forbidden region around $t = 0$ whereas a
nonvanishing cosmological constant allows a continuous transition between
the pre- and the post-Big Bang solutions without any initial singularity.
For these plots, we took $a_0 = 1$, $\alpha_0 = 0.5$, $A = 1$.}
}
\label{string}
\end{figure}

If we apply transformation (\ref {transfo1})-(\ref {transfo11}) to the
solution given by (\ref {sol11})-(\ref {sol13}), we obtain the following dual
solution:
\begin{eqnarray}
\bar{a}(t) & = & \displaystyle a_0^{-1}\,e^{-\,\alpha_0\,\tau(t)} 
\label{sol121}\\
\kappa\,\bar{\rho}_m(t) & = & \displaystyle A\,a_0^3\,
e^{3\,\alpha_0\,\tau(t)} \\
\bar{\phi}(t) & = & \displaystyle - 6\, ln (a_0)
-\,3\,\alpha_0\,\tau(t) + ln \left( C_1 \left[
cosh( \sqrt{3}\,\alpha_0\,\tau(t) ) - \delta_0 \right] \right)
\label{sol123}
\end{eqnarray}
where (\ref {constraints})-(\ref {case2}) are still valid.
For $a_0 = 1$ and $\delta_0 < 1$, the solution and its dual
can be joined smoothly without any singularity at $t = 0$. In order to
clarify this interesting feature, we plotted on the figure \ref {string}
the solution and its dual given by respectively (\ref {sol11})-(\ref {sol13})
and (\ref {sol121})-(\ref {sol123}) for $\delta_0 = 1$ and $\delta_0 =0.5$.
The figure \ref {string} clearly shows that it is impossible to allow a
smooth transition between the pre- and the post-Big Bang solutions without a
cosmological constant. Notice that the pre-Big Bang has been obtained by
making the time inversion $t \rightarrow - t$ in the dual solution.

\section{Non-minimally coupling with matter}

In this section, we consider quintessence models deriving 
from the following action
\begin{equation}
\displaystyle
S = \frac{1}{2\,\kappa}\,\int{
\left( R - \frac{1}{2}\,[\nabla\,\phi]^2 - V(\phi) \right)
\,\sqrt{-g\,}\,\,d^4 x}
\,+ \int{ f(\phi)\,L_m \,\sqrt{-g\,}\,\,d^4 x}
\label{action2}
\end{equation}
where the scalar field is assumed to be minimally coupled with the
gravitation and in interaction with the matter field. For
$\,f(\phi) = \,e^{2\,\phi}\,$, this action reduces to the string effective
action studied in the previous section but written in the Einstein frame,
i.e. after the following conformal
transformation $\,g_{\alpha\beta} \rightarrow e^{\phi}\,g_{\alpha\,\beta}\,$.

The case without the potential and with $\,f(\phi) = e^{2\,\phi}\,$
has already been studied in \cite{paper1}. If we add a potential
$V(\phi)$ and if we consider a more general coupling $f(\phi)$, we can
generalize the duality transformation found in \cite{paper1} as follows  
\begin{equation}
\left\{
\begin{tabular}{rlcl}
$G \,\,\rightarrow\,$ & $\bar{G}$ & $=$ & $q\,e^{- 2 \phi}\,G^{-1}$ \\ 
$\phi \,\,\rightarrow\,$ & $\bar{\phi}$ & $=$ & $\phi - ln\,q $ \\
$g_{00} \,\,\rightarrow\,$ & $\bar{g}_{00}$ & $=$ & $q\,g_{00}$ \\
$\rho_m \,\,\rightarrow\,$ & $f(\bar{\phi})\,\bar{\rho}_m$ & $=$ &
$q^{-1}\,f(\phi)\,\rho_m$  \\
$p_m \,\,\rightarrow\,$ & $f(\bar{\phi})\,\bar{p}_m$ & $=$ &
$-\,q^{-1}\,f(\phi)\,p_m$ \\
$V(\phi)\,\,\rightarrow\,$ & $\bar{V}(\bar{\phi})$ & $=$ &
$q^{-\,1}\,V(\phi)$
\end{tabular}
\right.
\label{transfo2}
\end{equation} 
where $q = e^{3 \phi}\,\mid\,det\,\tilde{G}\,\mid\,$ and where we have
assumed $\bar{f}(\bar{\phi}) = f(\bar{\phi})$ in order to keep the same 
coupling form after duality. As earlier, the equation of state
of the matter field remains the same under duality only for dust
($p_m = 0$).

The transformation on $V(\phi)$ given in (\ref {transfo2}) is necessary but
not sufficient. Indeed it does not imply that the dual and the initial
potentials are the same function of the scalar field, i.e.
$\bar{V}(\bar{\phi}) = V(\bar{\phi})$. It is straightforward to notice that
the potential form is maintained invariant only with an exponential potential
given by $\,V(\phi) = \Lambda\,e^\phi\,$. Then we have:
\begin{equation}
\displaystyle
\bar{V}(\bar{\phi}) = V(\bar{\phi}) = \Lambda\,e^{\bar{\phi}}\, =
\Lambda\,q^{-1}\,e^\phi = q^{-1}\,V(\phi)
\end{equation}

Furthermore, if we want the dual solution to be written in the cosmic
time, i.e. if we want $\,g_{00} \rightarrow \bar{g}_{00} = g_{00} = - 1\,$, 
it is necessary to add the following transformation on the 
timelike coordinate
\begin{equation}
\displaystyle
t \,\rightarrow\, \bar{t}(t) = \int{\sqrt{q(t)\,}\,dt}
\label{tftempo}
\end{equation}

When the background is given by a spatially flat FLRW metric ($g_{00} = -
1$), the field
equations derived from the action (\ref {action2}) with $\,V(\phi) =
\Lambda\,e^\phi\,$ and $f(\phi) = e^{2\,\phi}\,$ can been written as 
\begin{eqnarray}
\displaystyle
3\,\frac{\,a'^{\,2}}{a^2} & = & \displaystyle
\,\frac{1}{4}\, \phi'^{\,2} + \,\frac{1}{2}\,\,\Lambda\,\,e^\phi
+ \,e^{2\,\phi}\,\,\kappa\,\rho_m \label{eqn21} \\ 
\displaystyle
2\,\,\frac{\,a''}{a} + \frac{\,a'^{\,2}}{a^2} & = & \displaystyle 
- \frac{1}{4}\, \phi'^{\,2} + \,\frac{1}{2}\,\,\Lambda\,\,e^\phi
- \,e^{2\,\phi}\,\,\kappa\,p_m \label{eqn22} \\
\displaystyle
\phi'' + \,3\,\,\phi'\,\,\frac{\,a'}{a} & = & \displaystyle
- \,\Lambda\,\,e^\phi - \,e^{2\,\phi}\,\kappa\,\left[\,\rho_m
- 3\,p_m\,\right] \label{eqn23}
\end{eqnarray}
We obtain the following peculiar solution of this system:
\begin{equation}
\displaystyle
a(t) \propto \sqrt{\Lambda\,t^2\,/\,4} \hspace*{15mm}     
\phi(t) = -\,ln\left[\,\Lambda\,t^2\,/\,4\,\right] \hspace*{15mm}
p_m = \rho_m = 0
\label{sol2}
\end{equation}
\begin{figure}[h]
\begin{center}
\epsfig{figure=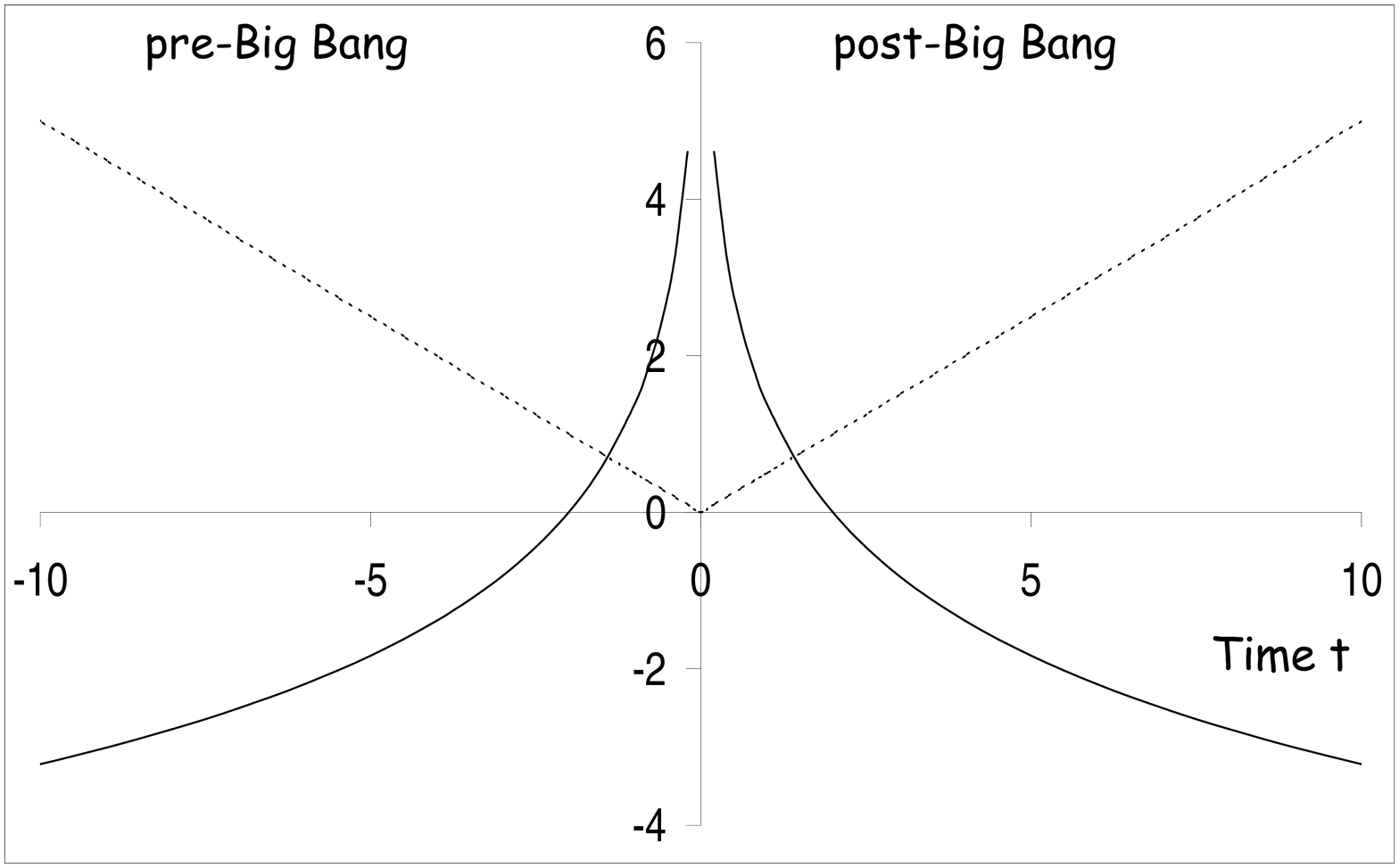,height=6cm,width=9cm}
\end{center}
\caption{
\small{The self-dual solution given by (\ref {sol2}) with $\Lambda = 1$ has
been used for the pre- and for the post-Big Bang phases. The dashed line
shows the scale factor behavior whereas the solid line represents the scalar
field.}
}
\label{fig2}
\end{figure}
This solution is ``self-dual'', i.e. it is invariant under the
transformations given by (\ref {transfo2})-(\ref {tftempo}). Figure
\ref {fig2} shows the behavior of this solution in the
pre- and post-Big Bang phases with $\Lambda = 1$. It has to be noted that
the initial singularity still exists at $t = 0$.

One way to avoid the additional timelike transformation given by (\ref
{tftempo}) on the dual solution is to work with the conformal time defined
by:
\begin{equation}
d\tau = \pm\, a^{-3}\,\,dt = \pm\, \bar{a}^{\,-3}\,\,d\bar{t} = d\,\bar{\tau}
\label{tau}
\end{equation}

If we apply the conformal transformation
$\,\,a\,\rightarrow\,a\,e^{-\phi/2}\,\,$ on the solution given by
(\ref {sol11})-(\ref {sol13}), we get another solution of equations
(\ref {eqn21})-(\ref {eqn23}) written in the conformal time:
\begin{eqnarray}
a(\tau) & = & \displaystyle \frac{a_0\,e^{-\,\alpha_0\,\tau\,/\,2}}
{\sqrt{C_1 \left[ cosh(\sqrt{3}\,\alpha_0\,\tau) - \delta_0 \right]}} 
\label{rien} \\
\kappa\,\rho(\tau) & = & \displaystyle
A\,a_0^{-\,3}\,e^{-\,3\,\alpha_0\,\tau} \\
\phi(\tau) & = & \displaystyle 3\,\alpha_0\,\tau + ln \left( C_1\,
\left[ cosh(\sqrt{3}\,\alpha_0\,\tau) - \delta_0 \right] \right)
\label{rien2}
\end{eqnarray}
with the constraints (\ref {constraints}). In this case, the conformal time
is related to the cosmic time by the relation (\ref {tau}). We have not been
able to find an analytical expression for the function $\tau(t)$. Using (\ref
{transfo2}), we obtain the following dual solution:
\begin{eqnarray}
\bar{a}(\tau) & = & \displaystyle \frac{a_0^2\,e^{\,\alpha_0\,\tau\,/\,2}}
{\sqrt{ C_1\,\left[ cosh(\sqrt{3}\,\alpha_0\,\tau) - \delta_0 \right]}}
\label{s1} \\
\kappa\,\bar{\rho}(\tau) & = & \displaystyle A\,a_0^3\,
e^{\,3\,\alpha_0\,\tau} \\
\bar{\phi}(\tau) & = & \displaystyle -\,3\,\alpha_0\,\tau - 6\,ln(a_0) + ln
\left( C_1\,
\left[ cosh(\sqrt{3}\,\alpha_0\,\tau) - \delta_0 \right] \right) \label{s2}
\end{eqnarray}
\begin{figure}[h]
\begin{center}
\epsfig{figure=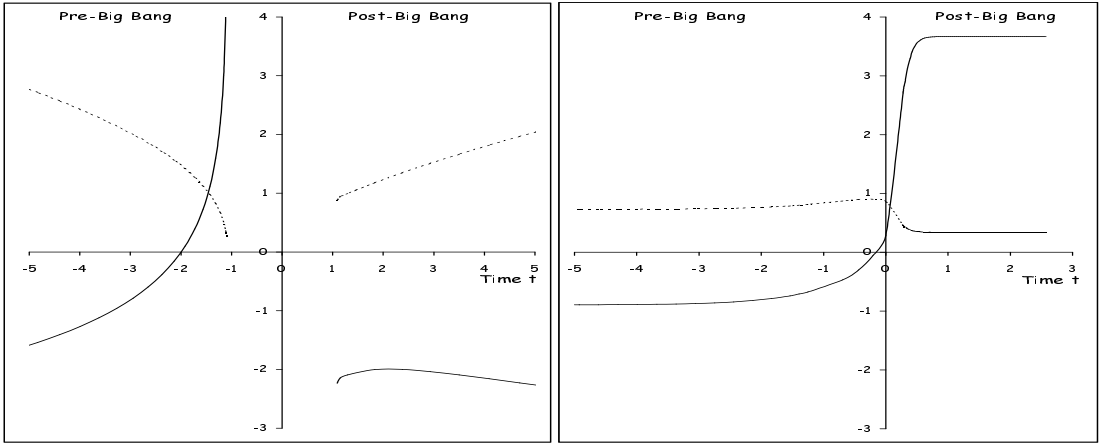,height=6cm,width=15cm}
\end{center}
\caption{
\small{The solutions given by (\ref {rien})-(\ref {rien2}) and
(\ref {s1})-(\ref {s2}) are plotted on this figure. The dashed
line gives the scale factor behavior whereas the solid line stands for the
scalar field. The case without the potential ($\delta_0 = 1$) stands on the
left graph whereas the one with a potential ($\delta_0 = 0.5$) is given on
the right. The use of a non-vanishing scalar potential allows a smooth
transition between the pre- and the post-Big Bang solutions. For these plots,
we took $a_0 = 1$, $\alpha_0 = 0.5$, $A = 1$.}
}
\label{string2}
\end{figure}

On the figure \ref {string2}, we plot this solution
(\ref {rien})-(\ref {rien2}) and its dual (\ref {s1})-(\ref {s2}) with
$a_0 = 1$. For comparison, the solution is represented with and without the
scalar potential. The relation between $t$ and $\tau$ has been numerically
computed. Again, for $a_0 = 1$ and in presence of the potential, the solution
and its dual can been smoothly connected, without any singularity.

\section{Minimally coupled scalar field}

Now we turn our attention to the quintessence models based on a minimally
coupled scalar field. These are of special interest because they are the
most invoqued in the framework of the supernovae observations. These models
are derived from the following action\footnote{The terms associated to the
scalar field usually appear in the action in another way, i.e.
$R - \,[\nabla\,\phi]^2 - 2\,V(\phi)$. We do not use these conventional
notations and keep the notations already used in \cite{paper1} so that no
factor $\sqrt{2}$ appears in the duality transformations.}:
\begin{equation}
\displaystyle
S = \frac{1}{2\,\kappa}\,\int{
\left[ R - \frac{1}{2}\,\,[\nabla\,\phi]^2 - V(\phi) \right]
\,\sqrt{-g}\,\,d^4 x} \,+ \int{L_m \,\sqrt{-g}\,\,d^4 x}
\label{action3}
\end{equation}
Taking into account what has been done in the previous section, it is easy to
show that the transformation given by
\begin{equation}
\begin{tabular}{rl}
$G \,\,\rightarrow\,$ & $\bar{G} = q\,e^{- 2\,\phi}\,G^{-1}$ \\ 
$\phi \,\,\rightarrow\,$ & $\bar{\phi} = \phi - ln\,q$ \\
$g_{00} \,\,\rightarrow\,$ & $ \bar{g}_{00} = q\,g_{00}$ \\
$V(\phi) = \Lambda\,e^\phi \,\,\rightarrow\,$ & $\bar{V}(\bar{\phi}) =
q^{-\,1}\,V(\phi) = \Lambda\,e^{\bar{\phi}}$ \\
$L_m \,\,\rightarrow\,$ & $\bar{L}_m = q^{-1}\,L_m$
\end{tabular}
\label{transfo3}
\end{equation}
with $q = e^{3\,\phi}\,\mid det\,G \mid\,$ does not modify the action (\ref
{action3}). The field equations deriving from action (\ref {action3})
with $\,V(\phi) = \Lambda\,e^\phi$ and $g_{00} = - 1$ can be written as:
\begin{eqnarray}
\displaystyle 3\,\frac{\,a'^{\,2}}{a^2} & = & \displaystyle
\frac{1}{4}\, \phi'^{\,2} + \,\,\frac{1}{2}\,\,\Lambda\,\,e^\phi
+ \,\,\kappa\,\rho_m \label{eqn31} \\
\displaystyle 2 \,\,\frac{\,a''}{a} + \frac{\,a'^{\,2}}{a^2} & = & 
- \frac{1}{4}\, \phi'^{\,2}
+ \,\,\frac{1}{2}\,\,\Lambda\,e^\phi - \kappa\,p_m \label{eqn32} \\
\displaystyle \phi'' + \,3\,\,\phi'\,\,\frac{\,a'}{a}
& = & \displaystyle -\,\Lambda\,e^\phi
\label{eqn33}
\end{eqnarray}
One can easily check that equations (\ref {eqn31})-(\ref {eqn33}) remain
invariant under the transformation given by (\ref {transfo3}) only if the
matter field changes as
\begin{equation}
\left\{
\begin{tabular}{rl}
$\rho_m \,\,\rightarrow\,$ & $\bar{\rho}_m = q^{-1}\,\rho_m$ \\
$p_m \,\,\rightarrow\,$ & $\bar{p}_m = q^{-1}\,[\, 2\,p_m - \rho_m\,]$
\end{tabular}
\right.
\label{transfo32}
\end{equation}
which means
\begin{equation}
\displaystyle
w_m = \frac{p_m}{\rho_m} \,\,\rightarrow\,\,\bar{w}_m =
\frac{\bar{p}_m}{\bar{\rho}_m} = 2\,w_m - 1
\end{equation}
In this case, the equation of state of the matter field is invariant under
the scale factor duality only for a stiff fluid ($w_m = 1$). On the
other hand, an initial dust fluid ($w_m = 0$) becomes a cosmological constant
($\bar{w}_m = -1$) after duality. Note that, as in the previous case, if we
want the dual solution to be written in the cosmic time, the duality
transformation given by (\ref {transfo3}) and (\ref {transfo32}) must be
followed by the transformation (\ref {tftempo}) on the timelike coordinate.

\section{Conclusions}

The scale factor duality is a symmetry property of the scalar-tensor theories
which has already been studied in the litterature. In this paper, we have
expanded those studies to quintessence models with a scalar potential. The
introduction of this potential has been dictated by the recent supernovae
observations which suggested a currently accelerating universe. Three kinds
of cosmological models have been considered in this paper: one with a
coupling between the
scalar field and gravitation, one allowing an interaction between the
scalar component and the matter field and the last one with a minimally
coupled quintessence field. In each case, the potential form which allows the
existence of a scale factor duality symmetry has been found.

In the framework of scalar-tensor theory where the scalar field is coupled
with gravitation, the potential form depends on the coupling, i.e. on the
gravitation theory. In the case of the effective string theory, only a theory
with constant scalar potential allows a scale factor duality. In the case of
a Brans-Dicke theory, the potential has to be linear whereas in a
scalar-tensor theory, the potential must have a power-law form.
Moreover, only the equation of state of a dust fluid is invariant under this
symmetry when the scalar field is not coupled with the matter field.

When there is no coupling between the scalar field and gravitation, the
potential must have an exponential form. In order to have an invariant
equation of state, the matter field must be dust, if interactions between
the scalar and matter components are allowed, and stiff, if the scalar field
is self-interacting.

Last but not least, we have also shown that in some cases, the scalar
potential allows a smooth transition between the pre- and the post-Big Bang
solutions which was not possible without such a potential. This is perhaps
the most important point. Indeed, till today, no satisfying model can explain
the transition between the pre- and the post-Big Bang phases. It would be
an interesting result if the potential necessary in the context of
quintessence would also permit to carry out this smooth transition: we have
shown that this is quite likely.
\\ \vspace*{1mm} \\ \noindent
{\large \bf{Acknowledgments}}
\vspace*{1mm} \\ \noindent
It is a pleasure to thank Sophie Pireaux and Jean-Marc G\'erard for helpful
comments on various aspects of the paper. The author is also grateful to
Jean-Pierre Swings for a careful reading of the manuscript. \\ \noindent
This work was supported in part by Belgian Interuniversity Attraction Pole
P4/05 as well as by a grant from ``Fonds National de la Recherche
Scientifique''.


\begin{thebibliography}{50}
\bibitem{quintessence}
B. Ratra and P. J. E. Peebles, {\em Phys. Rev. D} {\bf 37}, 3406 (1988);
R. R. Caldwell, R. Dave and P. J. Steinhardt, {\em Phys. Rev. Lett.}
{\bf 80}, 1582 (1998); P. J. Steinhardt, L. Wang and I. Zlatev, {\em Phys.
Rev. D} {\bf 59}, 123504 (1999); I. Zlatev, L. Wang and P. J. Steinhardt,
{\em Phys. Rev. Lett} {\bf 82}, 896 (1999); E. Di Pietro and J. Demaret,
to appear in Int. J. Mod. Phys. (gr-qc/9908071); M. S. Turner and M. White,
{\em Phys. Rev. D} {\bf 56}, 4439 (1997).
\bibitem{constant}
S. Weinberg, {\em Rev. Mod. Phys.} {\bf 61}, 1 (1989); 
V. Sahni and A. Starobinsky, {\em Int. J. Mod. Phys. D} {\bf 9}, 373 (2000).
\bibitem{duality}
M. Gasperini, hep-th/9907067 and the references therein;
An update collection of papers on string theory is available at 
\verb+http://www.to.infn.it/~gasperini/.+ 
\bibitem{pot}
M. Gasperini and G. Veneziano, {\em Astropart. Phys.} {\bf 1}, 317 (1993); 
M. Gasperini and G. Veneziano, {\em Phys. Lett. B} {\bf 277}, 256 (1992).
\bibitem{pot2}
K. A. Meissner and G. Veneziano, {\em Mod. Phys. Lett.} {\bf A6}, 3397
(1991);
\bibitem{paper1}
J. Demaret and E. Di Pietro, {\em Gen. Rel. Grav.} {\bf 31}, 323
(1999). E. Di Pietro and J. Demaret, {\em Int. J. Mod. Phys. D} {\bf 8},
349 (1999).
\bibitem{deritis}
R. de Ritis et al., hep-th/9907207.
\end{thebibliography}
\end{document}